\RequirePackage[2020-02-02]{latexrelease}
\documentclass[aps,prl,preprint,superscriptaddress,linenumbers]{revtex4}
\usepackage{graphicx}
\usepackage{dcolumn}
\usepackage{bm}
\usepackage{epstopdf}
\usepackage{csquotes}
\usepackage{amsmath}
\usepackage[utf8]{inputenc}

\begin{document}


\title{Fiber Loop Quantum Buffer for Photonic Qubits}

\author{Kim Fook Lee}
\affiliation{%
Center for Photonic Communication and Computing, Department of Electrical Engineering and Computer Science, Northwestern University, 2145 Sheridan Road, Evanston, IL 60208-3112, USA
}%
\author{Gamze Gul}
\affiliation{%
Center for Photonic Communication and Computing, Department of Electrical Engineering and Computer Science, Northwestern University, 2145 Sheridan Road, Evanston, IL 60208-3112, USA
}%
\author{Zhao Jim}
 \affiliation{%
 15 Presidential Way, Woburn, MA 01801, USA
 }%
 \author{Prem Kumar}
\affiliation{%
Center for Photonic Communication and Computing, Department of Electrical Engineering and Computer Science, Northwestern University, 2145 Sheridan Road, Evanston, IL 60208-3112, USA
}%
\email[]{kim.lee@northwestern.edu}



\date{\today}

\begin{abstract}

We report a fiber loop quantum buffer based on a low-loss 2$\times$2 switch and a unit delay made of a fiber delay line. We characterize the device by using a two-photon polarization entangled state in which one photon of the entangled photon pair is stored and retrieved at a repetition rate up to 78$\,\rm{kHz}$. The device, which enables integer multiples of a unit delay, can store the qubit state in a unit of fiber delay line up to 5.4$\,\rm{km}$ and the number of loop round-trips up to 3. Furthermore, we configure the device with other active elements to realize integer
multiplies and divider of a unit delay of a qubit.  The quantum state tomography is performed on the retrieved photon and its entangled photon. We obtain a state fidelity $>94\%$ with a maximum storage time of 52$\,\mu\rm{sec}$. To further characterize the storing and retrieving processes of the device, we perform entanglement-assisted quantum process tomography on the buffered qubit state. The process fidelity of the device is $>$ 0.98. Our result implies that the device preserves the superposition and entanglement of a qubit state from a two-photon polarization-entangled state.  This is a significant step towards facilitating applications in optical asynchronous transfer mode (ATM) based quantum networks.

\end{abstract}


\maketitle


\section{Introduction}

Fiber-based quantum communication and distributed quantum computing can provide secure internet over long distances with the help of quantum optical memories \cite{Gisin_RMP_2011,Dou_com_phys_2018,Wang_nphoton_2019,Tittel_nphoton_2009}. However, quantum memories with long storage times that utilize light-matter interactions are limited by optical bandwidth and low storage and retrieval efficiency \cite{Afzelius_NJP_2020,Wang_nphoton_2019}.
Quantum buffers made of fiber delay lines and free-space storage cavities can provide a reasonable storage time ($\mu\rm{sec}$) of the quantum states of light in different modes (polarization, time, and frequency)  without significant deterioration of storage efficiency \cite{XLi_OL_2005,Kwiat_SPIE_2021,Kwait_optica_2017,Kaneda_SA_2019}.

In general, variable optical buffers that provide long storage times ($\sim\mu\rm{sec}$) play an important role in optical
networks for packet synchronization, label processing, and contention management. Recently, optical packet switching has been introduced in quantum networks using classical-quantum data frames\cite{Alireza_PRR_2022}. Optical buffering can manage packet delay after label extraction and processing through classical data, routing the quantum data to the desired output port at the proper time without causing data congestion.
Quantum buffers based on fiber length have been studied and seem very promising for future quantum networking \cite{XLi_OL_2005, Park_OE_2019,Clemmen_OL_2018,Mendoza_Opa_2016}.
These buffers can be integrated with low-loss ($<$ 1$\,$dB) ultra-fast optical switches \cite{kflee_ptl_2021,kflee_ptl_2019} to facilitate the synchronization of packets in various platforms of packet-switching quantum networks.

Quantum buffers based on fiber delay lines are traveling buffers\cite{Zhong_2001} and fiber loop buffers\cite{Weber_1996}.
The former can allocate a fixed delay. The latter allows realizing of discrete variable delays.The combination of a few fixed delay lines and low-rate 2$\times$2 switches can provide active temporal source multiplexing for high-efficiency single photon generation\cite{Xiong_NC_2016,Zhang_OE_2017}. Fiber loop optical buffers enable data storage in a more compact footprint with respect to the traveling buffers. It is suitable for applications in asynchronous transfer mode (ATM) based quantum networks where packets are transmitted in cells of fixed delay.


When a photon propagates in a long fiber loop buffer,  its polarization qubit state  will experience the dephasing caused by the polarization modal dispersion (PMD) and the loss due to the absorption/scattering process in the fiber reservoir. The fiber reservoir is the medium where a polarization qubit interacts/exchanges energy with the environment. The absorption and re-emission of a photon state in the fiber reservoir will decohere a quantum state. It is essential to characterize the density matrix of the buffered photon and its polarization-entangled photon by using quantum state tomography. In the fiber loop buffer, the qubit storing and retrieving processes are initialized by a 2$\times$2 switch. These switching processes can impose loss and decoherence on the buffered qubit. Quantum process tomography ($\chi-$tomography)\cite{Chuang_JMO_1997} can provide a precise characterization of a quantum process or operation that acts on the buffered qubit. The process tomography has been used to characterize the qubit state retrieved  from a free-space storage cavity\cite{Kwiat_SPIE_2021}.

In this paper, we construct a fiber loop buffer by using a 2$\times$2 switch and a fiber delay line. The delay line is connected to one of the input ports and one of the output ports of the switch to form a fiber loop. One photon of the polarization-entangled photon pair can be directed to propagate in a fiber loop. To store and retrieve a photonic qubit in the loop for N round-trips, we need to apply different rates of the radio frequency (RF) electrical signal (ON and OFF) to the 2$\times$2 Switch. The photon storage time is then given by N $\times$ the length of the fiber loop/ speed of light in the fiber. When we operate the switch by using the RF electrical signal (ON and OFF) with the rate of $>$ 50kHz, the buffered photon is leaking through the buffer before it is retrieved from the buffer.  We resolve the leaking issue of the 2$\times$2 switch by using a different output port of the switch to form a loop. We first perform the quantum state tomograph of the polarization-entangled photon pair where one photon of the pair passes through the fiber loop buffer for N=1, 2, and 3. We then perform quantum process tomography on the buffered photon. We conclude that the fiber loop buffer can preserve the superposition and entanglement of the buffered photonic qubit even though the 2$\times$2 switch induces the loss and other decoherence processes such as the amplitude/phase damping process.

\section{Results}

In our two-qubit system, the photon pair is initially prepared in the entangled state $|\psi_{\circ}\rangle=\frac{1}{\sqrt{2}}[|H_s H_i\rangle + |V_s V_i\rangle]$.
One photon of the entangled pair is sent to the fiber loop buffer with the fiber loop length of L.
We use a counter-propagating scheme (CPS) \cite{Kflee_ol_2006,Sua_ol_2014} to generate a two-photon polarization entangled state $|\psi_{\circ}\rangle$ through a four-wave mixing process in a 300$\,$m of dispersion-shifted fiber.
The signal and idler photons are separated by the DWDMs (dense wavelength-division multiplexing). The signal photon is sent to the polarization analyzer $\rm{PA_{s}}$. The idler is sent to the loop buffer which consists of a 2$\times$2 switch (insertion loss $<$1.0$\,$dB) and a fiber loop made of a single-mode optical fiber as shown in Fig.1.
We can select the idler to pass a fiber loop for N round-trips.
The idler is then retrieved and directed to the polarization analyzer $\rm{PA_{i}}$.
The first pair of the quarter-wave plate (QWP) and half-wave plate (HWP) is used to compensate birefringence of the fiber. The second pair of QWP and HWP is used to implement quantum state tomography (QST) and quantum process tomography. We use two single photon detectors (NuCrypt CPDS-4) operated at 50$\,$MHz for measuring the coincidence and accidental.

\begin{figure}[htbp]
\centering
\includegraphics[scale=0.5]{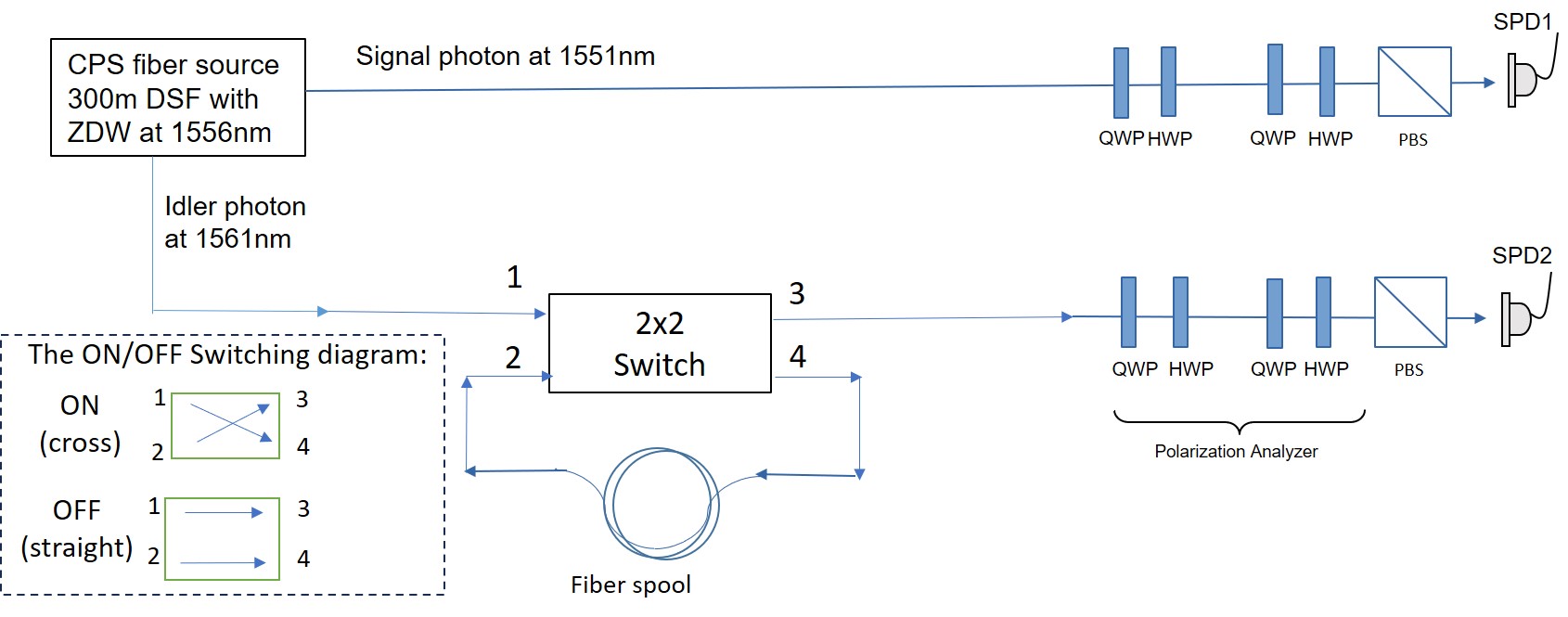}
\caption{\textbf{Experiment setup for demonstrating the fiber loop buffer made of one low-loss 2$\times$2 switch and a fiber delay line by using a fiber-based polarization-entangled photon pair. The polarization analyzers are used for projecting 16 measurements of the $|\psi_{\circ}\rangle = (|H_{s} H_{i}\rangle + |V_{s}V_{i}\rangle)/\sqrt{2}$. Inset: The ON/OFF switching diagram of the 2$\times$2 switch when it is turned ON/OFF by the RF. The insertion loss (including the connector) of the switch is about 1.2dB (ON: cross) and 1.0dB (OFF: straight).}}
\label{Figure1}
\end{figure}

We characterize the prepared state $|\psi_{\circ}\rangle = (|H_{s} H_{i}\rangle + |V_{s}V_{i}\rangle)/\sqrt{2}$ by using the standard method of QST \cite{James_pra_2001,White_pra_2001,Munro_pra_2001}. There are a total of 16 settings for the HWPs and QWPs in the $\rm{PA_{s}}$ and $\rm{PA_{i}}$.
For each setting, we measure the coincidence counts (CC) and accidentals (AC) with an integration time of 2 seconds.
Using the maximum likelihood estimation, we reconstruct the $4 \times 4$ density matrix of the final state $\rho$ in the HV basis.
We construct the density matrix $\rho$ based on the true coincidence counts i.e. the accidental coincidence is subtracted. In other words, the Raman photon and loss-induced accidental do not contribute to this study.
As for the entanglement-assisted quantum process tomography (EAPT), the signal photon is used as a triggered photon and the idler is sent to the buffer. The coincidence and accidental are measured for the post-process states (storing and retrieving processes) of the idler photon and the signal's triggered detector. The advantage of the EAPT is we can use the same measurement data of the 16 settings for constructing the $\chi-$tomography of the buffered idler photon\cite{Altepeter_PRL_2003}.

\begin{figure}[htbp]
\centering
\includegraphics[scale=0.5]{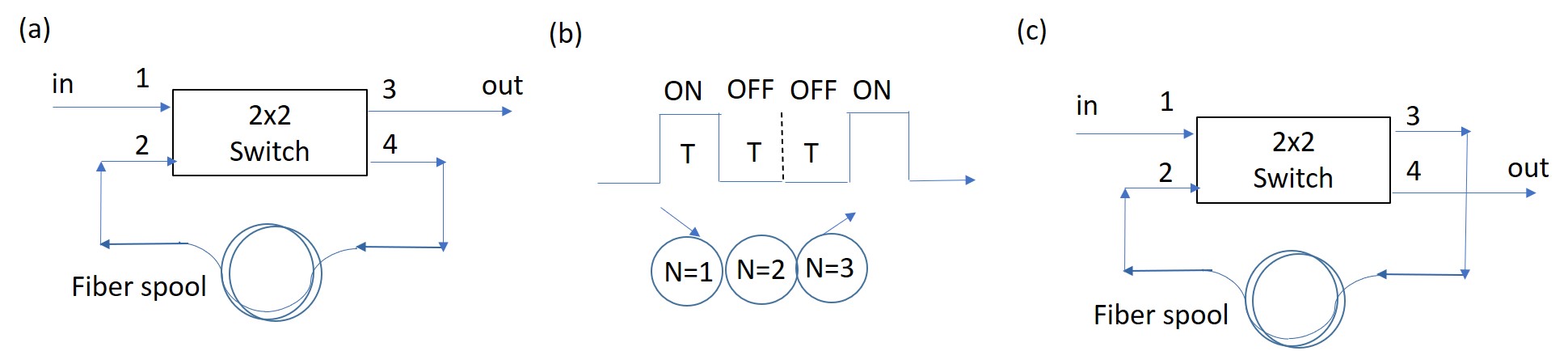}
\caption{\textbf{a) The fiber loop is connected to input port 2 and output port 4 of the switch for the RF rate $<$ 50kHz. b) Showing the N = 3 round-trips by using RF's ON/OFF pattern : T duration of the ON state and 2T duration of the OFF state. c) We connect the fiber loop to input port 2 and output port 3 to optimize the switch for completely turn ON and OFF the buffer for the RF rate $>$ 50kHz.}}
\label{Figure2}
\end{figure}

We configure a buffer by using one 2$\times$2 switch as shown in Fig. 2(a). We send the photon to the buffer through input 1 and measure the retrieved photon at output 3. We connect input port 2 and output port 4 with a fiber loop of the length $L$. We use the RF pulse to turn ON and OFF the 2$\times$2 switch. The ON/OFF switching diagram of the input/output ports is shown in the inset of Fig.1. We use the switch to store the photon to the fiber loop (1$\rightarrow$4) and then to call out the photon from the fiber loop (2$\rightarrow$ 3).
When the idler is injected into the loop, the idler is buffered with the duration of $\rm{N}\times\frac{L}{c/n_{r}}$, where N is the number of round-trips, $L$ is the length of the fiber, and $n_{r}$ is the refractive index of the fiber.
The duration of the RF's ON state (which is the start of the first round-trip) is defined by the length of the fiber loop. While the duration of the RF's OFF state, which is the integer multiplies of the duration of the ON state, is corresponding to the $N-1$ number of round-trips.
For example, we can implement the N = 3 round-trips by using the T duration of the ON state and 2$\times$T duration of the OFF state as illustrated in Fig.2(b).
We can control the N round-trips by adjusting the durations of the ON state and the OFF state of the RF. The duration T is limited by the repetition rate of the switch. The switch has rise and fall times of about 100$\,$ns but its repetition rate is up to 100$\,$kHz. We are able to operate the switch at the repetition rate up to 78$\,$kHz, which is corresponding to the shortest T duration time of $\frac{1}{2}\times\frac{1}{\rm{78kHz}}=6.4\mu\rm{sec}$ and a fiber loop length of 1.3$\,$km.

We first test the switch by applying the DC voltage to the 2$\times$2 switch and connecting a 2 meters single-mode fiber (SMF) patch cord to the input port 2 and output port 4 of the switch. The buffer acts like a traveling buffer (N = 1) with a fixed delay line of 2 meters (buffer time of 9.7ns).
We perform quantum state tomograph for the entangled photon state and obtain the reconstructed density matrix $\rho$ with the fidelity F $>$ 0.97.
We also perform entanglement-assisted quantum process tomography for the buffered idler. We obtain the fidelity of the $\chi-$ tomography of about 0.98. The measured $\chi-$ tomography indicates that the identity process ($I_{\circ}$) is 0.98. Note that the identity process ($I_{\circ}$) = 1.00 corresponds to no decoherence process induced by the 2$\times$2 switch on the qubit.
We then change the delay line of the traveling buffer by using a fiber with a length of 5.4km (buffer time $\sim 26\mu\rm{sec}$).  We obtain the reconstructed density matrix $\rho$ with the fidelity F $>$ 0.95. We also perform entanglement-assisted quantum process tomography of the buffer. We obtain the $\chi-$ fidelity of about 0.98. The drop in the state fidelity from 0.98 to 0.95 is due to the PMD effect on the buffered qubit. The PMD can degrade the polarization extinction ratio of the buffered qubit.
These results indicate that the traveling buffer, where the 2$\times$2 switch operated at the DC voltage, can preserve the qubit state and entanglement of the two-photon polarization entangled state.

We then apply the RF to the 2$\times$2 switch so that we can implement the fiber loop buffer for $\rm{N} > 1.0$. We apply the RF at the repetition rate of 19kHz, 25.6kHz, and 34.5kHz for the fiber loop with the length of 5.4km, 4.0km, and 3.0km, respectively, the idler photon propagates through the fiber loop twice or N = 2 round trips. The total buffer length (time) for the fiber loop of 5.4km, 4.0km, and 3.0km at 19kHz, 25.6kHz, and 34.5kHz is $2 \times 5.4\rm{km} = 10.8\rm{km}$ ($52.0\mu\rm{sec})$, $2 \times 4.0\rm{km} = 8.0\rm{km}$ ($39.0\mu\rm{sec})$, and $2 \times 3.0\rm{km} = 6.0\rm{km}$ ($29.0\mu\rm{sec})$, respectively.  We perform quantum state tomography and process tomography of the buffered photon and its entangled photon at each RF repetition rate. For the maximum buffer length (time) of $2 \times 5.4\rm{km} = 10.8\rm{km}$ ($52.0\mu\rm{sec}$), we obtain the reconstructed density matrix $\rho$ with the fidelity F = 0.94 and the $\chi-$ tomography with the fidelity of 0.95 as shown in Fig. 3(a) .
The fidelities of the state tomography and $\chi-$ tomography for N = 1, 2, and 3 are shown in Table I.

\begin{figure}[htbp]
\centering
\includegraphics[scale=0.5]{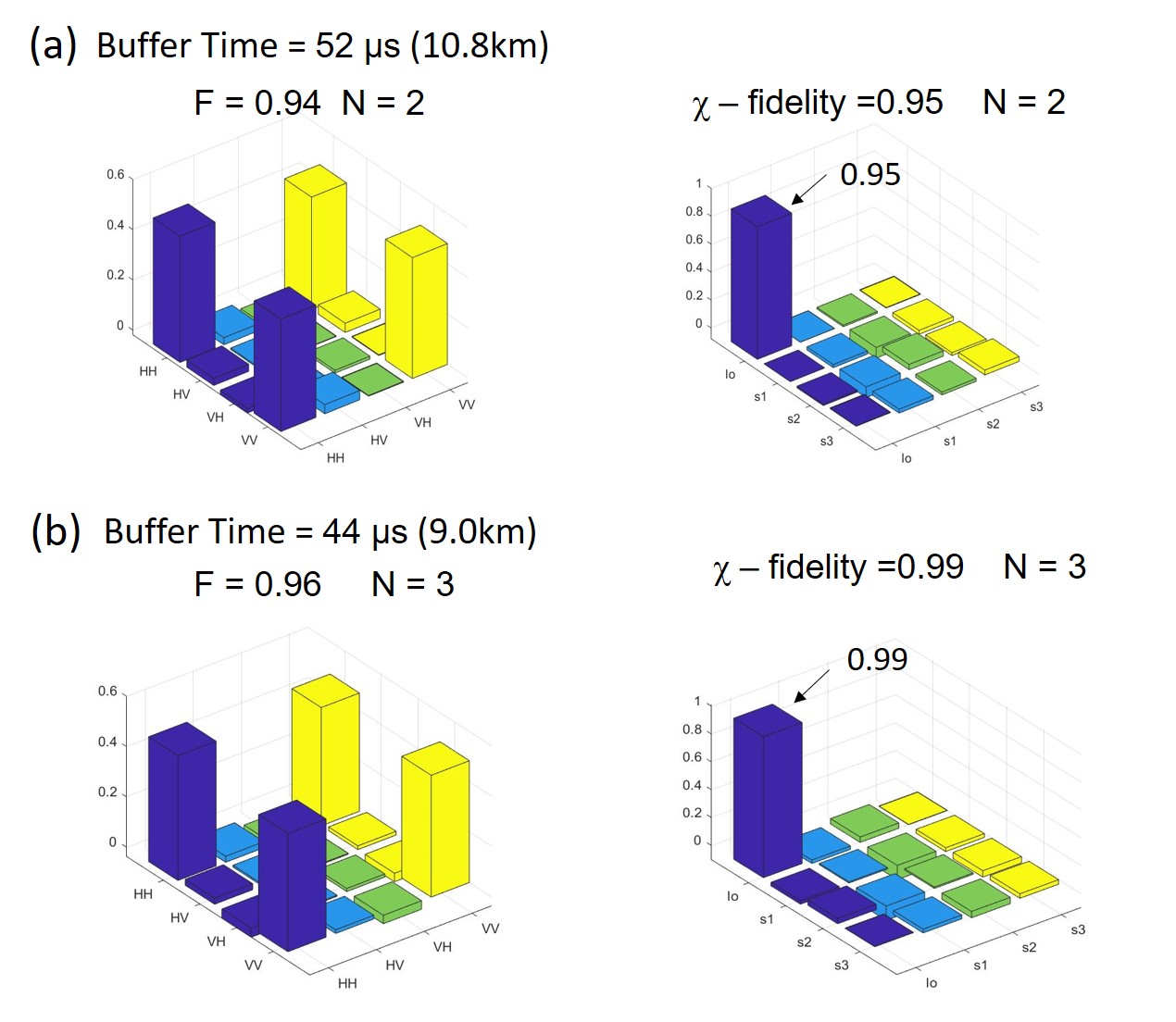}
\caption{\textbf{a) Left: The reconstructed density matrix of $|\psi_{\circ}\rangle = (|H_{s} H_{i}\rangle + |V_{s}V_{i}\rangle)/\sqrt{2}$ for N = 2.0 with the buffer length of 2$\times$5.4km = 10.8km (buffer time =52$\mu$s) and the state fidelity F = 0.94. Right: The obtained  $\chi-$ tomography with the $\chi-$fidelity = 0.95. b) Left: The reconstructed density matrix of $|\psi_{\circ}\rangle = (|H_{s} H_{i}\rangle + |V_{s}V_{i}\rangle)/\sqrt{2}$ for N = 3 with the buffer length of 3 $\times$3.0km = 9.0km (buffer time = 44$\mu$s) and the state fidelity F = 0.96. Right: The obtained  $\chi-$ tomography with the $\chi-$ fidelity = 0.99. Notation: $I_{\circ}$ is Identity Operator. The $s_1$, $s_2$, and $s_3$ are Pauli Operators for x, y, and z, respectively. }}
\label{Figure3}
\end{figure}

As for N=3, we apply the RF with the ON and OFF electrical signal such that ON for 14.7$\mu\rm{sec}$ and OFF for $2\times14.7\mu\rm{sec}$ at the rate of $\frac{1}{3\times14.7\mu\rm{sec}}$ = 22.7kHz. The idler photon propagates through the fiber loop for N = 3 round trips. The total buffer length is 3 $\times$ 3.0km = 9.0km and the total buffer time is about $3\times14.7\mu\rm{sec} = 44 \mu\rm{sec}$. We then perform quantum state tomograph of the quantum buffer. The reconstructed density matrix $\rho$ with the fidelity F $>$ 0.96 is shown in Fig.3(b: Left). We also perform entanglement-assisted quantum process tomograph of the buffered idler photon. We obtain the $\chi-$fidelity of about 0.99 as shown in Fig.3(b: Right).
The obtained state and process fidelities for N = 3.0 are similar to N = 2.0 with the same fiber loop length of 3.0km.  This indicates that the fiber loop buffer can preserve the qubit state and entanglement at different RF's ON/OFF patterns.

\begin{table} [h!]
\centering
\resizebox{\textwidth}{!}{
\begin{tabular}{|c|c|c|c|c|c|}
\hline
N(pass)-fiber loop & Buffer length (time) & Total Insertion loss & Quantum State & Fidelity(F) & $\chi-$Fidelity \\
  \hline
  1 & $\rm{5.4km(26\mu s)} $& $\rm{<1.2dB(cross)\times2+0.2dB/km\times5.4km=3.48dB}$ & $|\psi_{\circ}\rangle$ &0.95 & 0.98 \\
  2 & $\rm{2\times5.4km=10.8km(52\mu s)}$ &$\rm{<1.2dB(cross)\times2+1.0dB(straight)+0.2dB/km\times10.8km=5.56dB}$ & $|\psi_{\circ}\rangle$ & 0.94 & 0.95 \\
  $2^{\star}$ & $\rm{2\times4.0km=8.0km(39\mu s)}$ &$ \rm{<1.2dB(cross)\times2+1.0dB(straight)+0.15dB/km\times8.0km=4.60dB}$ & $|\psi_{\circ}\rangle$ & 0.95 & 0.99 \\
  2 & $\rm{2\times3.0km=6.0km(29\mu s)}$ &$ \rm{<1.2dB(cross)\times2+1.0dB(straight)+0.2dB/km\times6.0km=4.60dB}$ & $|\psi_{\circ}\rangle$ & 0.95 & 0.98 \\
  2 & $\rm{2\times1.82km=3.66km(17.9\mu s)}$ &$ \rm{<1.2dB(cross)\times2+1.0dB(straight)+0.2dB/km\times3.66km=4.13dB}$ & $|\psi_{\circ}\rangle$ & 0.95 & 0.99 \\
  2 & $\rm{2\times1.30km=2.60km(12.7\mu s)}$ &$ \rm{<1.2dB(cross)\times2+1.0dB(straight)+0.2dB/km\times2.60km=3.92dB}$ & $|\psi_{\circ}\rangle$ & 0.95 & 0.98 \\
  3 & $\rm{3\times3.0km=9.0km(44\mu s)}$ &$\rm{ <1.2dB(cross)\times2+1.0dB\times2(straight)+0.2dB/km\times9.0km=6.20dB}$ & $|\psi_{\circ}\rangle$ & 0.96 & 0.99 \\
  \hline
\end{tabular}}
\caption{Showing the figure of merit of the fiber loop buffer for N = 1, 2, and 3. $\star$ is where the SMF28 ULL (ultra-low loss) is used for the fiber loop.  Notation: 'cross' is RF ON and 'straight' is RF OFF}
    \label {table:1}
    \end{table}

When we apply the RF at the rate of 55.8kHz for N =2 with the fiber loop length of 1.85km, the idler photon in the fiber loop leaks from port 2 to the output port 3 before it is retrieved by the second RF's ON signal. We then prevent the leaking by adjusting the $\pi-$ voltage ($V_{pi}$) setting of the 2$\times$2 switch and reconnecting the fiber loop to port 2 and port 3 as shown in Fig.2(c). In this new configuration, the 2$\times$2 switch can operate without leaking at the RF rate up to 78kHz. This is corresponding to N = 2 round-trips with a fiber loop length of 1.3km. For both RF at 55.8kHz and 78kHz, the state and process fidelities are $>$0.95 and $>$0.98, respectively, as shown in Table I. The reconstructed density matrix $\rho$ and $\chi -$ tomography for the RF at 78kHz is shown in Fig.4.

\begin{figure}[htbp]
\centering
\includegraphics[scale=0.5]{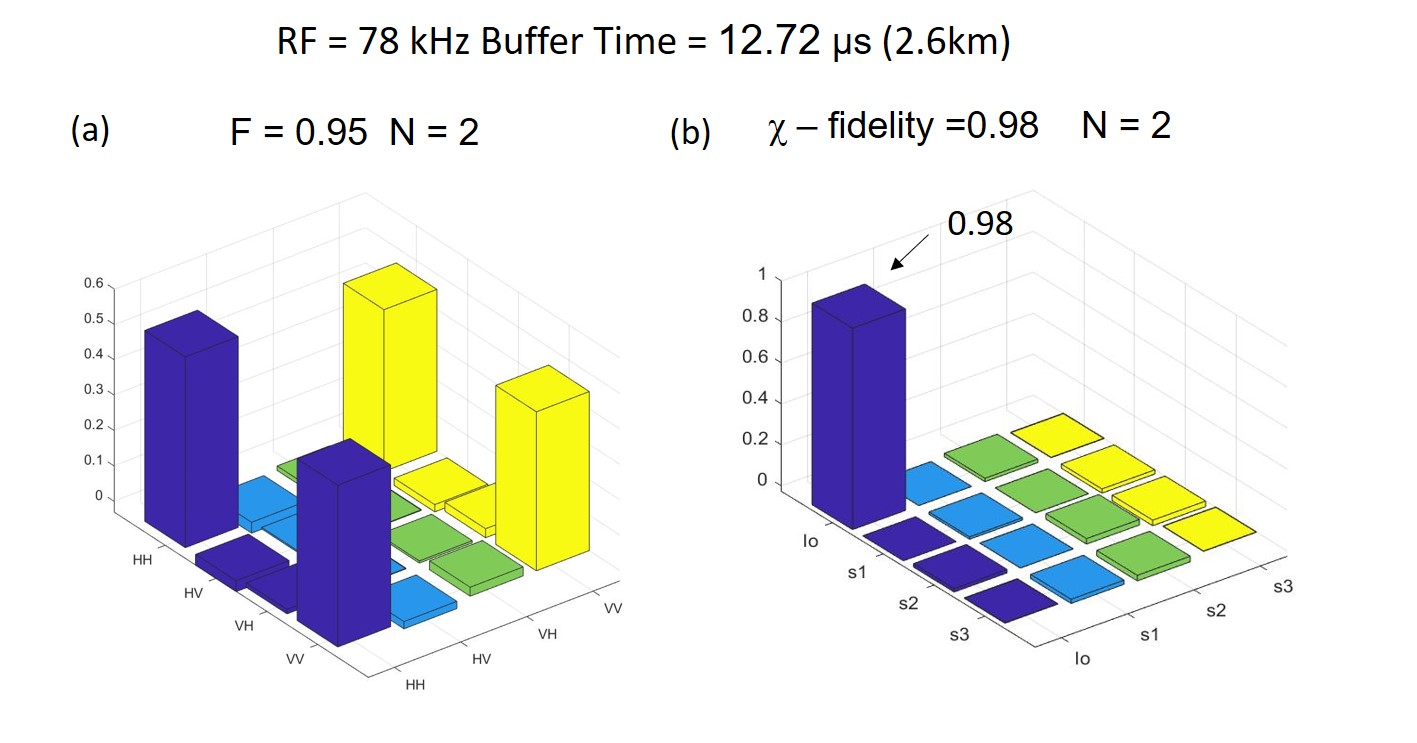}
\caption{\textbf{a) The reconstructed density matrix of $|\psi_{\circ}\rangle = (|H_{s} H_{i}\rangle + |V_{s}V_{i}\rangle)/\sqrt{2}$ with the RF at 78kHz and N = 2 with the buffer length of 2 $\times$ 1.3km = 2.6km (buffer time = 12.72$\mu$s) and the state fidelity F = 0.95. b) The obtained $\chi -$ tomography with the RF at 78kHz and N = 2 and the $\chi-$fidelity = 0.98}}
\label{Figure4}
\end{figure}

It is important to show that the fiber loop can be integrated with other switches to provide more selective buffer times for optical packet synchronization and contention management. Here, we insert two units of low-loss switches ($<$0.01dB) in the loop configuration as shown in Fig.5a. We use the single-mode ultra-low loss fiber (SMF28 ULL) from Corning as the fiber delay lines. The transmission loss is 0.15dB/km for the c-band telecom wavelength. The 1.0km fiber delay line is inserted parallel to the 4.0km fiber delay line between the two 1$\times$2 switches. The design of using a 1.0km fiber length is to implement the principle of the integer divider (D) of a unit delay which is made of the 4.0km ($19.57\mu\rm{sec}$) fiber length (buffer time). To illustrate this principle, we apply the RF at the repetition rate of 25.6kHz to the 2$\times$2 switch and then select the fiber loop of 4.0km by activating port 7 and port 8 of the 1$\times$2 switches. This scenario, which is similar to the above experiments, provides the integer multiples (N) of a unit delay i.e. N = 2 round-trips of 4.0km  with a total buffer time of about 2$\times19.57\mu\rm{sec} = 39.1\mu\rm{sec}$. When the 1.0km fiber delay line is selected by activating port 5 and port 6 of the 1$\times$2 switches, we have a buffer time of about 4.9$\mu\rm{sec}$ which is exactly 39.1$\mu\rm{sec}$ divided by 8. This design provides the integer multiples (N = 2) of a unit delay and integer divider (D=4) of a unit delay. The  1$\times$2 switches are synchronized and operated at the repetition rate of 1.0Hz which is designed to be slower than the 25.6kHz of the 2$\times$2 switch. We obtain three different buffer times i.e. 0.0sec, 4.9$\mu\rm{sec}$, and 39.1$\mu\rm{sec}$. We also observe the buffered photon at the buffer time of about 5 $\times 4.9\mu\rm{sec}$ with negligible single counts. This is due to the ends of the ON duration and the OFF duration of the 2$\times$2 switch as illustrated in Fig.5b. We then perform quantum state and process tomography on the buffered idler photon and its entangled signal photon for 3 buffer times. For the buffer times  of 0.0s, $4.9\mu\rm{s}$, and $39.1\mu\rm{s}$, we obtain the state fidelity ($\chi-\rm{fidelity}$)  of 0.97 (0.99), 0.97(0.99), and 0.95(0.99), respectively.

\begin{figure}[htbp]
\centering
\includegraphics[scale=0.5]{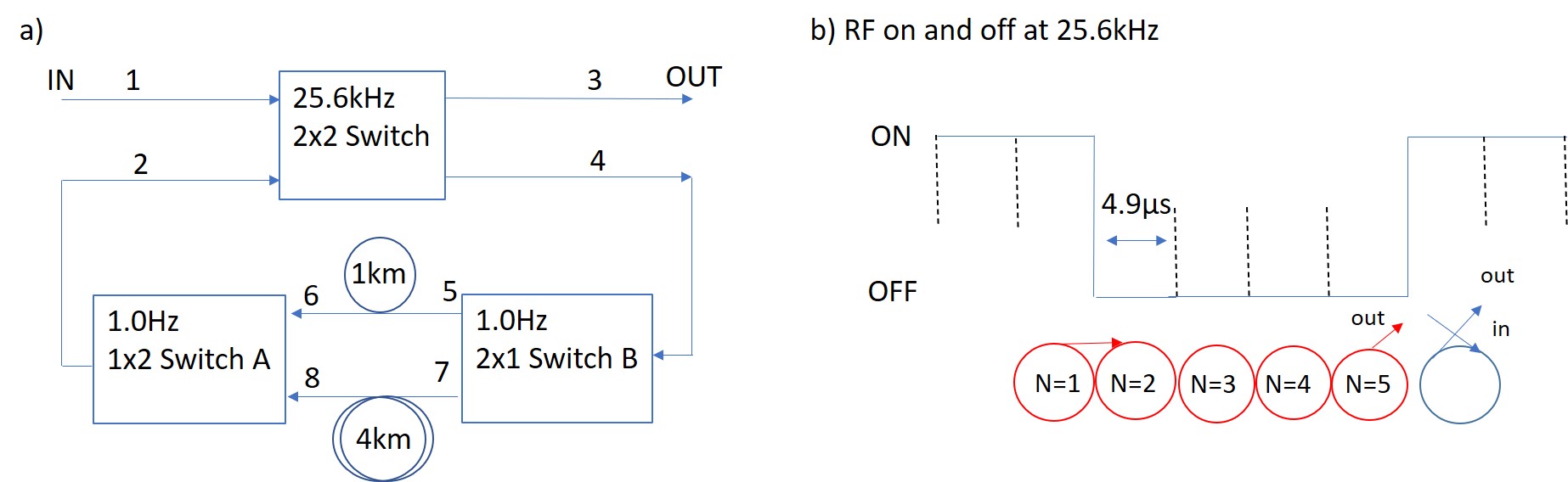}
\caption{\textbf{a) The setup for implementing 3 buffer times by using a 2$\times$2 switch and two units of 1$\times$2 switch. Port 1 (3) is the input (output) port of the buffer. The fiber delay lines of 1km and 4.0km are the SMF28 ULL. b) Showing the  fourth buffer time i.e. $5\times4.9\mu\rm{s}$ inside the 1.0km fiber loop. The idler is buffered for N = 5 in 1.0km fiber loop with negligible single counts at the output port 3 because of the insertion loss of the 2$\times$2 switch.}}
\label{Figure5}
\end{figure}

\section{Discussion}

We have demonstrated that our fiber loop buffer can preserve entanglement by storing and retrieving one photon from the polarization-entangled photon pairs up to $\rm{52\mu\rm{sec}}$.  The polarization-entangled idler photon consists of the superposition state $|H\rangle_{i}+|V\rangle_{i}$. The 2$\times$2 switch can preserve the superposition of the idler qubit state because it has negligible polarization-dependent loss (PDL).
Our fiber loop design enables discrete storage times i.e. integer multiples of a unit delay. The design also allows easy implementation of quantum state tomography and quantum process tomography on the photon state.
The buffer can preserve the quantum state fidelity $>$ 0.94 for the storage time ranging from $\rm{26\mu s}$ to $\rm{52\mu s}$ and the number of round-trips (N) up to 3. Even though the switching and retrieving processes of the 2$\times$2 switch dominate the insertion loss of the buffer as shown in Table I, we obtain the $\chi-$ fidelity in the range from 0.95 to 0.99. This indicates that the switch induces negligible decoherence processes such as amplitude/phase damping process, bit-flip, bit-phase flip, and phase-flip.
Furthermore, we have demonstrated that the fiber loop buffer together with other active elements such as 1$\times$2 switches enable us to realize integer multiplies (N) and divider (D) of a unit delay of a photon state with the preservation of the quantum state and $\chi-$ fidelities of about  $>$0.95 and $>$ 0.99, respectively.
The developed device could be used as a variable quantum buffer for controlling packet delays and allowing flexible contention management.
Our fiber loop buffer design can be integrated with the chip scale buffers\cite{Takesure_ncomm_2013,Wang_2022} which can provide high-resolution buffering of sub-100ps time scale and facilitate the synchronization of the photons to achieve a fully
programmable integrated quantum network.

\section{Data availability}
The data that supports the findings of this study are available from the corresponding author upon reasonable request.

\begin{acknowledgments}

\end{acknowledgments}


\begin{thebibliography}{27}

\section{References}

\bibitem{Gisin_RMP_2011}
Sangouard, Nicolas, Simon, Christoph, de Riedmatten, Hugues \& Gisin, Nicolas. Quantum repeaters based on atomic ensembles and linear optics. {\em Rev. Mod. Phys.} \textbf{83}, 33-80 (2011).

\bibitem{Dou_com_phys_2018}
Dou, J.P., Yang, A.L., Du, M.Y. et al.  A broadband DLCZ quantum memory in room-temperature atoms. {\em Commun Phys} \textbf{1}, 55 (2018).

\bibitem{Wang_nphoton_2019}
Wang, Y., Li, J., Zhang, S., Su, K., Zhou, Y., Liao, K., Du, S., Han, Y., \& Zhu, S.L. Efficient quantum memory for single-photon polarization qubits. {\em Nat. Photon.} \textbf{13}, 346-351 (2019).

\bibitem{Tittel_nphoton_2009}
Lvovsky, A., Sanders, B. \& Tittel,W. Optical quantum memory. {\em Nat. Photon.} \textbf{3}, 706-714 (2009).

\bibitem{Afzelius_NJP_2020}
Holzapfel, A., Etesse, J., Kaczmarek, K. T., Tiranov, A., Gisin, N., \& Afzelius, M. Optical storage for 0.53 seconds in a solid-state atomic frequency comb memory using dynamical decoupling. {\em  New J. Phys.} \textbf{22}, 063009 (2020).

\bibitem{XLi_OL_2005}
Li, Xi., Voss, P. L., Chen, J., Sharping, J. E. \& Kumar, P. Storage and long-distance distribution of telecommunications-band polarization entanglement generated in an optical fiber. {\em Opt. Lett.} \textbf{30}, 1201-1203 (2005).


\bibitem{Kwiat_SPIE_2021}
 Arnold, N. T., Victora, M., Goggin, M. E.  \& Kwiat, P. G. All-
optical ultrawide-bandwidth quantum buffer. {\em Proc. SPIE, Photonics for
Quantum } \textbf{11844}, 118440Y (2021).

\bibitem{Kwait_optica_2017}
Kaneda, F., Xu, F., Joseph, C. \& Kwiat, P. G. Quantum-memory-assisted multi-photon generation for efficient quantum information processing. {\em Optica} \textbf{4}, 1034-1037 (2017).

\bibitem{Kaneda_SA_2019}
Kaneda, F.\& Kwiat, P. G. High-efficiency single-photon generation via large-scale active
time multiplexing. {\em Sci. Adv.} \textbf{5} eaaw8586 (2019)

\bibitem{Alireza_PRR_2022}
DiAdamo, Stephen and Qi, Bing and Miller, Glen and Kompella, Ramana \& Shabani, Alireza. Packet switching in quantum networks: A path to the quantum Internet. {\em Phys. Rev. Res.} \textbf{4}, 043064 (2022).

\bibitem{Park_OE_2019}
Eunjoo Lee, Sang Min Lee \& Hee Su Park. Relative time multiplexing of heralded telecom-band single-photon sources using switchable optical fiber delays. {\em Opt. Express} \textbf{27}, 24545-24555 (2019).

\bibitem{Clemmen_OL_2018}
S. Clemmen, Alessandro Farsi, Sven Ramelow \& Alexander L. Gaeta. All-optically tunable buffer for single photons. {\em Opt. Lett.} \textbf{43}, 2138-2141 (2018).

\bibitem{Mendoza_Opa_2016}
G. J. Mendoza, R. Santagati, J. Munns, E. Hemsley, M. Piekarek, E. Martin-Lopez, Graham D. Marshall, Damien Bonneau, Mark G. Thompson \& Jeremy L. OBrien. Active temporal and spatial multiplexing of photons. {\em Optica} \textbf{3}, 127-132 (2016).


\bibitem{kflee_ptl_2021}
Lee,K. F., Moraw, P. M., Reilly, D. R.  \& Kanter, G. S. Pulse Retiming for Improved Switching Rates in Low-Noise Cross-Phase-Modulation-Based Fiber Switches. {\em IEEE Photonics Technology Letters} \textbf{33} 51-54 (2021) doi: 10.1109/LPT.2020.3044256.


\bibitem{kflee_ptl_2019}
Lee, K. F.  \& Kanter, G. S. Low-Loss High-Speed C-Band Fiber-Optic Switch Suitable for Quantum Signals. {\em IEEE Photonics Technology Letters} \textbf{31} 705-708 (2019) doi: 10.1109/LPT.2019.2905593


\bibitem{Zhong_2001}
Zhong, W. D. \& Tucker, R. S.  A new wavelength-routed photonic packet buffer combining traveling delay lines with delay-line loops. {\em J. Lightwave Technol.} \textbf{19} 1085-1092 (2001).


\bibitem{Weber_1996}
Langenhorst, R., Eiselt, M., Pieper, W., Grosskopf, G.,  Ludwig, R., Kuller, L.,  Dietrich, E. \&H. G. Weber. Fiber loop optical buffer. {\em J. Lightwave Technol.} \textbf{14} 324-335 (1996).

\bibitem{Xiong_NC_2016}
Xiong, C. et al. Active temporal multiplexing of
indistinguishable heralded single photons. {\em Nat. Commun.} \textbf{7} 10853 (2016). doi: 10.1038/
ncomms10853.

\bibitem{Zhang_OE_2017}
Zhang, X., Lee, Y. H., Bell, B. A., Leong, P. H. W., Rudolph, T., Eggleton, B. J. \&  Xiong, C. Indistinguishable heralded single photon generation via relative temporal multiplexing of two sources. {\em Opt. Express} \textbf{25} 26067-26075 (2017).


\bibitem{Chuang_JMO_1997}
Chuang, I. S. \& Nielsen, M. A. Prescription for experimental determination of the dynamics of a quantum black box. {\em J. Modern Optics} \textbf{44} 2455-2467 (1997).



\bibitem{Kflee_ol_2006}
Lee, K. F., Chen, J., Liang, C. , Li, X., Voss, P. L. \& Kumar, P. Generation of high-purity telecom-band entangled photon pairs in dispersion-shifted fiber. {\em Opt. Lett.}
 \textbf{31}, 1905-1907 (2006).

\bibitem{Sua_ol_2014}
Sua, Y. M., Malowicki, J. \& Lee, K. F. Quantum correlation of fiber-based telecom-band photon pairs through standard loss and random media. {\em Opt. lett.} \textbf{39}, 4808-4811 (2014).

\bibitem{James_pra_2001}
James, Daniel F. V., Kwiat, Paul G.,  Munro, William J. \& White, Andrew G. Measurement of qubits. {\em Phys. Rev. A.} \textbf{64}, 052312 (2001).

\bibitem{White_pra_2001}
White, A. G., James, D. F., Munro, W. J. \& Kwiat, P. G. Exploring Hilbert space: Accurate characterization of quantum information. {\em Phys. Rev. A.} \textbf{65}, 012301 (2001).

\bibitem{Munro_pra_2001}
Munro, W. J., James, D. F., White, A. G. \& Kwiat, P. G. Maximizing the entanglement of two mixed qubits. {\em Phys. Rev. A.} \textbf{64}, 030302 (2001).


\bibitem{Altepeter_PRL_2003}
Altepeter, J. B., Branning, D., Jeffrey,E. R., Wei, T. C., Kwiat, P. G., Thew, R., O'Brien, J. L., Nielsen, M. \& White, A. G. Ancilla-assisted quantum process tomography.  {\em Phys. Rev. Lett.} \textbf{90} 193601 (2003).


\bibitem{Takesure_ncomm_2013}
Takesue, H., Matsuda, N., Kuramochi, E. et al. An on-chip coupled resonator optical waveguide single-photon buffer. {\em Nat. Commun.} \textbf{4}, 2725 (2013). https://doi.org/10.1038/ncomms3725

\bibitem{Wang_2022}
Wang, X. \& Mookherjea, S. Feasibility of chipscale integration of single-photon switched digital loop buffer. {\em Chip} \textbf{1} 100028 (2022). https://doi.org/10.1016/j.chip.2022.100028

\end{thebibliography}

\newcommand{\noopsort}[1]{} \newcommand{\printfirst}[2]{#1}
  \newcommand{\singleletter}[1]{#1} \newcommand{\switchargs}[2]{#2#1}

\section{Acknowledgments}

This work was supported in part by the DOE (DE-SC0020537)

\section{Author Contributions}
K.F.L and P.K conceived the research.  K.F.L analyzed the data, wrote the paper and prepared the manuscript.

\section{Competing Interests}
The authors declare no competing interests.

\end{document}